# Compton Scattering of a Twisted Light

Mazen Nairat, and David Voelz

Physics Department, Balqa`a Applied University,

Salat, Jordan

Electrical and Computer Engineering Department,

New Mexico State University, USA

## ABSTRACT

The variation of photonic orbital angular momentum at Compton scattering is characterized. We determine scattering matrix of a twisted light based on the fundamental conservation of orbital angular momenta. Numerical values for two different twisted light modes: Laguerre Gaussian and Bessel Gaussian, are generated and illustrated. Our analysis indicate that states of photonic orbital angular momentum are highly changeable at wide angle scattering but more consistent at small angle scattering.

1. Introduction:

Twisted light carries well-defined photonic orbital angular momentum (POAM). Its wave front characterizes specific azimuthal phase according to POAM state [1]. Both photonic spin angular momentum and POAM are separate components of total light angular momentum.

Compton scattering describes the change in linear momentum at elastic collisions between a photon and an electron. The well-known shift in a scatter wave number reports certain change in the linear momentum as well as in the energy of scattering photons [2]. However, a variation in POAM had never been mentioned.

Several studies have been recently conducted to describe changes in POAM at Compton scattering in ultra-relativistic considerations [3, 4]. A non-relativistic framework has been implemented in the density matrix theory to inform the variation of POAM at Compton scattering [5].

Our study briefly analyzes a change in POAM of twisted light in Compton scattering by evaluating the associated scattering matrix at semi classical physics framework. It illustrate the possibility for POAM to vary through scattering by atomic electrons. It also emphasizes the conservation of total angular momentum regards exchange POAM into massive electrons.

We evaluate in the next section analytical expression for scattering matrix of Compton scattering for a twisted light. The expression is valid for any axial symmetric beam carries well defined POAM. In the following section, two different beams for instance have been examined. Our numerical calculations are then analyzed and discussed. The study is finally summarized.

2. Conservation of OAM in Compton scattering:

The schematic diagram of Compton scattering in Fig. 1 identifies the scattered photonic wave number (*k*) as well as POAM by primed parameters (*l*), while the associated wave number and orbital angular momentum of the recoil electron is indicated by $k_e$ and $m_e$, respectively. The scattering angle ($\theta$) is defined as the angle between the incident beam and the scattering direction.

Our analysis bases on evaluating semi classically the scattering matrix of Compton scattering to illustrate the fundamental of exchange the orbital angular momentum between twisted light wave and massive particle.

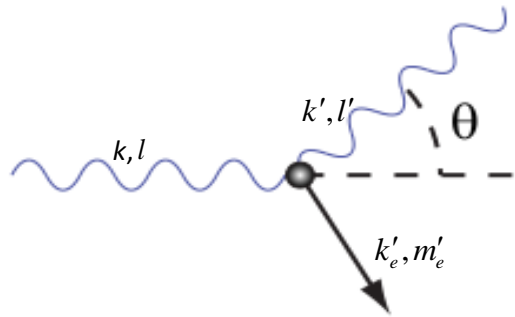

Fig.1: Schematic of Compton scattering with associated parameters.

Compton scattering matrix is determined as $S = \left\langle f' \left| e^{i(\vec{k}'-\vec{k}) \cdot \vec{r}} \right| f \right\rangle$ where: $f'$, $f$ represent scattered and initial state of the system, $\vec{k}$, $\vec{k}'$ are incident and scattered wave vectors, and $\vec{r}$ is a position vector[6]. The cylindrical coordinates are used in such a way the optical path is considered along z-axis.

Both spatial electronic and photonic wave functions, Φ and Ψ respectively, are considered. Their conjugates are indicated by superscript stars as well, Compton scattering matrix could be presented as

$$S = \langle \Phi^*(\vec{r}')\Psi^*(\vec{r}') | e^{i(\vec{k}'-\vec{k}).\vec{r}} | \Phi(\vec{r})\Psi(\vec{r}) \rangle \tag{1}$$

Normalized wave functions in cylindrical coordinates are in used due to azimuthal symmetry of twisted light. Generally; electronic and photonic wave functions could be expressed as:

$\Phi(\vec{r}) = P_m(\rho, z)e^{-im\phi}$ , $\Psi(\vec{r}) = R_l(\rho, z)e^{-il\phi}$ where m,l are parameters indicate electronic state and POAM state, respectfully. Therefore, Eq (1) is written explicitly as:

$$S = \langle P_{m'}^*(\rho', z')e^{i(m'+l')\phi}R_{l'}^*(\rho', z') | e^{i(\vec{k}'-\vec{k}).\vec{r}} | P_m(\rho, z)e^{-i(m+l)\phi}R_l(\rho, z) \rangle \tag{2}$$

The azimuthal symmetry of the spatial waves profiles splits exponential of azimuthal parameters out as separate term as follow:

$$S = \langle P_{m'}^*(\rho', z')R_{l'}^*(\rho', z') | e^{i(\vec{k}'-\vec{k}).\vec{r}} | P_m(\rho, z)R_l(\rho, z) \rangle \langle e^{i(\Delta m + \Delta l)\phi} \rangle \tag{3}$$

where $\Delta m = m' - m$ and $\Delta l = l' - l$. The last angle bracket of Eq (3) is evaluated directly using Dirac delta function:

$$\langle e^{i(\Delta m + \Delta l)\phi} \rangle = \delta(\Delta m + \Delta l) \tag{4}$$

Eq (4) reads that a decrease in POAM must be equivalent to an increase of electronic one: $\Delta m = -\Delta l$. Consequently, Compton scattering matrix is not terminated as long as orbital angular momentum exchanges between the photon and the electron, it must be conserved through scattering.

Evaluating Compton scattering matrix requires resolving the wave functions as well as middle exponential term. Electronic radial wave function $P_m(\rho, z)$ is separable function, it is generally given by:

$$P_m(\rho, z) = J_m(\kappa_e \rho)e^{ik_e z} \tag{5}$$

where $J_m$ is Bessel function, $\kappa_e$ and $k_e$ are transverse and longitudal electronic pointing vector, respectively. On the other hand, wave vector of collimated twisted light in general composites of two components: orbital and longitudal [7]. It can be basically presented as

$$\vec{k} = \kappa\hat{\phi} + k\hat{z} \tag{6}$$

Hence, middle term in the scattering matrix, the exponential term in Eq.3, is simplified as follow:

$$\exp[(\vec{k}' - \vec{k}).\vec{r}] = \exp[(k' - k)z] \tag{7}$$

Eqs (4, 5 and 7) are implemented to evaluate scattering matrix as follow

$$S = \langle J_{m'}^*(\kappa'_e\rho')R_l^*(\rho',z') | J_m(\kappa_e\rho)R_l(\rho,z) \rangle \, \delta(\Delta k_e + \Delta k) \tag{8}$$

Dirac delta term indicates that longitudinal component of the pointing vector, which represents the linear momentum, is also exchanged between the electron and the photon. An increased in linear momentum of the electron is equivalent to a decrease in linear momentum of the photon. It is definitely consistent with well-known Compton`s formula [2]. Conservation of both linear and orbital momentum emphasizes conservation of total momenta for twisted photons in Compton Scattering.

Our objective is to focus on the change of photonic orbital angular momentum in Compton scattering. It leads evaluating scattering matrix regardless initial OAM state. It is computed for a relative OAM`s change in zero order state "untwisted mode" i.e.:

$$S = \langle J_{\Delta l}^*(\kappa'\rho)R_{\Delta l}^*(\rho,z') | J_0(\kappa\rho)R_0(\rho,z) \rangle \tag{9}$$

Conditions of Eq.3 and Eq.7 have been used. Indeed, It has been assumed there is no change in radial parameter ($\rho$) through scattering, on the mean while axial parameter (*z*) depends only on the scattering angle ($\theta$).Thus, the scattering matrix is evaluated at certain axial distance.

3. Results and discussion

Scattering matrix of Eq.9 is computed for two particular twisted light beams: *Bessel Gauss* and *Laguerre Guassian (LG)*. The spatial wave function of LG is given by [8]:

$$R_l^{LG}(\rho,z) = \frac{C_p^l}{w(z)}\left[\frac{\sqrt{2}\rho}{w(z)}\right]^{|l|} \exp\left(-\frac{\rho^2}{w(z)^2}\right) L_p^{|l|}\left(\frac{2\rho^2}{w(z)^2}\right) \exp\left[\frac{ik\rho^2 z}{2(z^2+z_R^2)} + iG_q^{|l|}(z)\right] \tag{10}$$

where $C_p^l$ is a normalization constant, *p* is radial index, *w(z)* is the beam width at distance *z* and given by $w(z) = w_o\sqrt{1+(z/z_R)^2}$ ,$z_R$ is Rayleigh range which is defined as: $z_R = \pi w_o^2/\lambda$, $L_p^{|l|}(\bullet)$ is associated LeGender polynomial, and $G_p^{|l|}(z) = -i(2p+|l|+1)\tan^{-1}(z/z_R)$ is Gouy phase. It should be noted that scattered axial parameter, z`, varies only through the scattered angle ($\theta$). Our computations bases on fixed z value.

Eqs.9 and 10 are used to evaluate scattering matrix of an x-ray of 1 nm wavelength and 1µm waist width through wide scattering range $-\pi/2 < \theta < \pi/2$. It is assumed that there is no change in Gouy phase through scattering.

The computed elements that associate for a certain change (Δ*l*) in photonic OAM are determined and plotted. Fig.2 illustrates Compton scattering matrix vs scattering angle for several changes in POAM. The curves are generated by computing normalized scattering matrix on wide angle range for four successive states change in POAM.

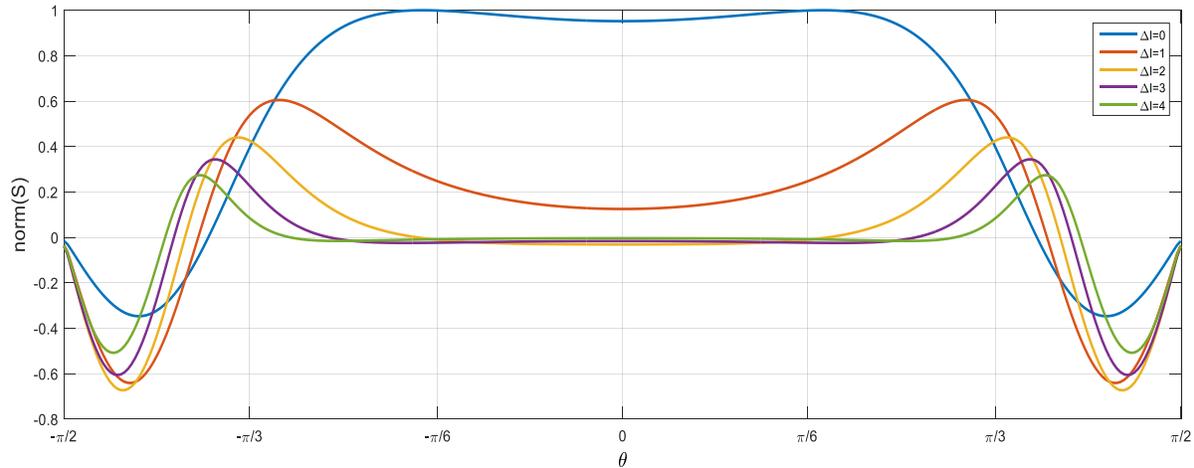

Fig.2: Normalized Compton matrices vs scattering angle for LG beam at certain changes in POAM

The normalized scattering elements shown in Fig.2 are very small for narrow range scattering. The normalized scattering provides the probability of exchanging POAM which is minimum at small scattering angle. However, a change by just one order state could happen with low possibility as the curve of Δ*l*=1 shows.

High possibilities of changing POAM states through scattering are represented by the peaks in Fig2. However, scattering matrix elements are relatively small values for higher change in POAM, It could be interpreted as a low possibility for a big change in POAM at Compton Scattering.

Negative-value elements in Fig.2 indicate flipping of OAM sates up, they inform that scattering in wide angles causes flipping azimuthal phase up. Consequently, POAM states are twisted "oppositely" at wide angles Compton-scattering. However; Scattering matrices are terminated at right angle scattering due to the minimize differential cross section of Compton scattering [9].

Scattering Matrix is also computed for large changes in photonic OAM states and plotted in Fig.3

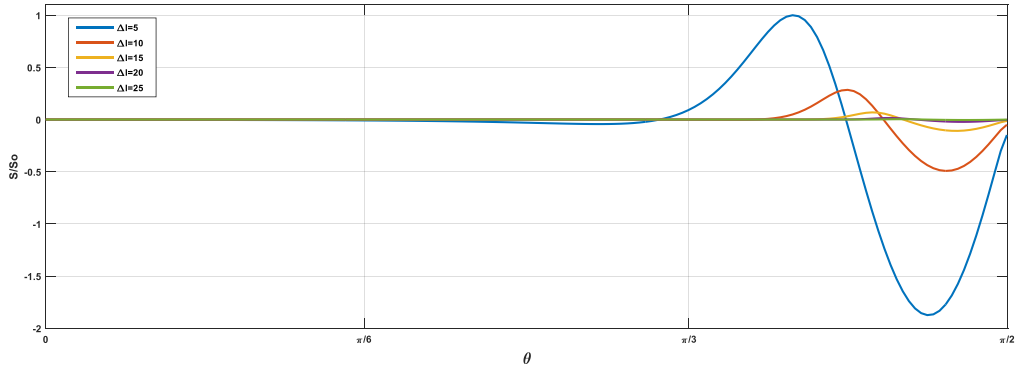

Fig.3: Normalized Compton matrices vs scattering angle for LG beam at big changes in photonic OAM

Fig.3 confirms the low possibilities of scattering twisted photons with high changes in their POAM. Indeed, it indicates Compton scattering at wide angles is more likely to occur with opposite twisted orientation. Dominant negative peaks at Fig. 3 represent a tendency of large changes in photonic OAM to be in opposite orientation.

The other twisted light beam that has been investigated is Bessel Gaussian beam which has spatial wave given by [10]

$$R_l^{GB}(\rho,z) = A\frac{w_o}{w(z)}\exp\left[i\left(k-\frac{\kappa^2}{2k}\right)z - i\phi(z)\right]J_l\left[\frac{\kappa\rho}{(1+iz/z_R)}\right]\exp\left[\left(\frac{-1}{w(z)^2}+\frac{ik}{2R(z)}\right)\left(\rho^2+\frac{\kappa^2}{k^2}z^2\right)\right]$$

(11)

where, A is normalization constant, $\phi(z) = \tan^{-1}(z/z_R)$ is the associated Gouy phase, and $R(z) = z\sqrt{1+(z_R/z)^2}$ is the radius of curvature.

Eq. 11 has been implanted in, Eq. 9 to compute the corresponding scattering matrix, same parameters: a nm wavelength of an x-ray source, a μm beam waist and non-varying Gouy phase.

The associated scattering matrix is evaluated for first order changes in orbital angular as shown in Fig. 4

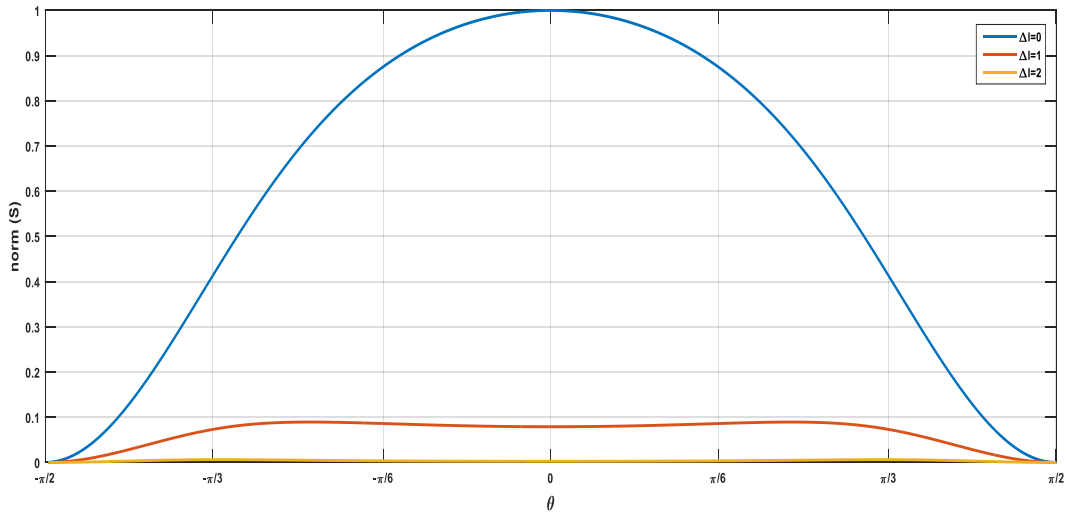

Fig.4: Normalized Compton matrices vs scattering angle for BG beam at few changes in photonic OAM

Scattering matrices of Bessel Gauss beam that illustrate in Fig. 4 change very slowly, relative high values peaks around scattering angle of $\pi/3$ compare with small values at small scattering angles, $|\theta| \leq \pi/6$. It likes to Laguerre Gaussian, Photonic OAM of Bessel Gauss beam is likely to invariant at small angle scattering, they just tends to vary in wide scattering, however; extreme minimum scattering elements is at right angle scattering where classical cross section is minimum.

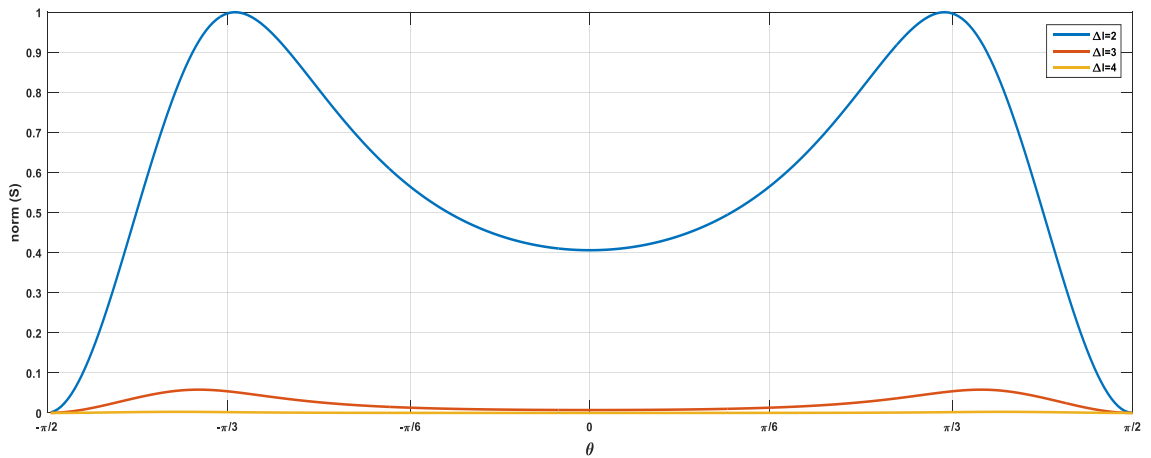

Fig.5: Normalized Compton matrices vs scattering angle for BG beam at certain changes in POAM

Scattering matrix is also computed for higher order changes in POAM of Bessel Gaussian mode as illustrated in Fig.5. The order is changed by varying associated azimuthal parameter to 2, 3 and 4, respectively.

Fig 5 shows emphasis the high possibility to exchange OAM at wide scattering compare with small angles scattering. POAM are more consistent in Compton scattering by small angles, although extreme lowest values at right angle scattering.

### Conclusion

This study analyzes conservation of POAM at Compton scattering. We evaluated scattering matrix for twisted light that scatter by atomic electrons. Our analysis indicate POAM of a twisted light could be changed through Compton scattering. Specific numerical values are determined for Compton scattering of Laguerre Gaussian and Bessel Gaussian light beams. It has been reported that POAM states are invariant at forward scattering as well as small angle scattering, they mainly vary wide angle scattering.

### Acknowledgments

Authors thank Roman Höllwieser, The affiliate assistant professor at New Mexico State University, for checking our calculations as well as our results.

**References**

[1] L. Allen, M. W. Beijersbergen, R. J. C. Spreeuw, and J. P. Woerdman "*Orbital angular momentum of light and the transformation of Laguerre-Gaussian laser modes*" Phys. Rev. A 45, 8185

[2] Compton, Arthur H. (May 1923). "*A Quantum Theory of the Scattering of X-Rays by Light Elements*". Phys Rev 21 (5): 483–502. Bibcode:1923PhRv...21..483C

[3] Ivanov, I.P. and V.G.Serbo" Scattering of twisted particles: Extension to wave packets and orbital helicity", Phys Rev A, 84 (3): 033804-9.

[4] U.D. Jentschura and V.G. Serbo, "Generation of High-Energy Photons with Large Orbital Angular Momentum by Compton Backscattering", Phys. Rev. Lett. 106, 013001.

[5] S. Stock, A. Surzhykov, S. Fritzsche, and D. Seipt, "*Compton scattering of twisted light: angular distribution and polarization of scattered photons*", AIP, PACS: 42.50.Tx, 03.65.Nk, 2015.

[6] B. S. Davis, L. Kaplan, and J. H. McGuire, "*On Exchange of Orbital Angular Momentum Between Twisted Photons and Atomic Electrons*", J. Opt. **15** 109501, 2013

[7] M. PADGETT and L. ALLEN, "Light with a twist in its tail", Cont. Phys., **41** 5, 2000

[8] L. Allen and M. Babiker "Spin-orbit coupling in free-space Laguerre-Gaussian light beams", Phys. Rev. A 53, R2937

[9] Klein, O; Nishina, Y. "Über die Streuung von Strahlung durch freie Elektronen nach der neuen relativistischen Quantendynamik von Dirac". *Z. Phys.* **52** (11-12):1929

[10] Gori F., Guattari G. and Padovani C., "Bessel-Gauss Beams", *Opt. Commun.*, **64**, 491, (1987)